\documentclass{optica-article}

\journal{opticajournal} 

\articletype{Research Article}

\usepackage{graphicx} 
\usepackage{caption}
\usepackage{subcaption}
\usepackage{braket}
\usepackage{makecell}
\usepackage[leftcaption]{sidecap}
\usepackage{lineno}
\usepackage{graphicx} 
\graphicspath{ {images/} }

\newcommand{\murm}{%
  \ifmmode
    \mathchoice
        {\hbox{\normalsize\textmu}}
        {\hbox{\normalsize\textmu}}
        {\hbox{\scriptsize\textmu}}
        {\hbox{\tiny\textmu}}%
  \else
    \textmu
  \fi
}

\begin{document}

\title{Absolute Length Sensing in a Long-Baseline, High-Finesse Optical Cavity}

\author{Todd Kozlowski$^{1,}$*, Henry~Fr\"adrich, Aaron D. Spector}

\address{Deutsches Elektronen-Synchrotron DESY, 22603 Hamburg, Germany}

\email{\authormark{*}todd.kozlowski@desy.de} 


\begin{abstract*} 
The relative phase between two lasers in transmission of an optical cavity can be used to continuously measure its absolute length with sub-micron precision. The first laser is kept on resonance with the cavity, while a second laser is phase-locked to the first with a frequency separation equal to an integer multiple of the cavity's initial free spectral range. As the free spectral range frequency changes due to cavity length changes, the second laser detunes slightly from resonance and gains an additional phase offset in transmission of the cavity. The cavity length changes can be calibrated in terms of this phase offset. This technique is applied to a high-finesse optical cavity with a length of 123 meters, transforming it into a strainmeter with nanostrain sensitivity to long-term and transient seismic events. We report absolute length changes associated with anthropogenic noise, a distant earthquake, and the diurnal and semidiurnal earth tides.
\end{abstract*}

\section{Introduction} \label{sec:introduction}
Long-baseline Fabry-Perot cavities are commonly employed to measure the distance between two mirrors. In interferometric gravitational wave detectors, kilometer-scale cavities are used to amplify the phase change of light resulting from relative length changes between two arms of a Michelson interferometer \cite{Abbott2009,Accadia2012}. Techniques can also be employed using double modulation of resonant sidebands to determine the absolute length of an optical cavity to high precision \cite{Araya1999,Staley2015}. This method has also been used to perform long-term observations of Earth strain and record large strain steps caused by local earthquakes \cite{TAKEMOTO2004, Takamori2014}. In this paper we report on a technique for obtaining the absolute length of a Fabry-Perot cavity using a second resonant laser field and performing a phase-sensitive heterodyne readout in transmission. We applied this method to continuously measure the precise length of the 123-meter long ALPS\,II Regeneration Cavity \cite{Kozlowski2024}. This yielded not only a recording of long-term earth tidal strain, but also sensitive measurements of transient, anthropogenic noise produced by nearby cultural events.

\section{Theory} \label{sec:theory}
The amplitude and phase of laser light in transmission of a Fabry-Perot cavity is modified as a function of the laser's frequency relative to cavity resonance, $\Delta \nu$, called here detuning. 
The cavity complex transmission coefficient $\mathcal{T}(\Delta \nu)$ is the ratio of the transmitted and incident fields. For high-finesse cavities,
\begin{equation}
    \mathcal{T}(\Delta \nu) \equiv \frac{E_{\rm trans}}{E_{\rm in} } \approx \frac{\sqrt{T_{\rm in}}\sqrt{T_{\rm out}}}{\displaystyle\frac{A}{2} - 2\pi i\displaystyle\frac{\Delta\nu}{f_0}},
    \label{Eq:E_t}
\end{equation}
where $T_{\rm in}$ and $T_{\rm out}$ are the input and output cavity mirror transmissivities, $A$ is the total round-trip attenuation in the cavity including excess optical losses $l$ ($A\equiv T_{\rm in} + T_{\rm out} + l$), and $f_0$ is the cavity free-spectral-range (FSR). The FSR is the frequency spacing between subsequent fundamental cavity resonances, and is given as $f_0 = c/(2L)$ in vacuum, where $L$ is the cavity length. The absolute length of the cavity is then determined by measuring $f_0$. The phase component of Eq. \ref{Eq:E_t} can be expressed as,
\begin{equation}
    \phi(\Delta \nu) = \arctan{\left(\frac{2\pi \Delta \nu /f_0}{A/2}\right)} = \arctan{\left( \frac{2 \mathcal{F}\Delta\nu}{f_0}\right)}
\end{equation}
where in the second equality, $\mathcal{F} = 2\pi / A$ is the cavity finesse (valid for $A \ll 1$).

\begin{figure}
    \centering
    \includegraphics[width=0.8\linewidth]{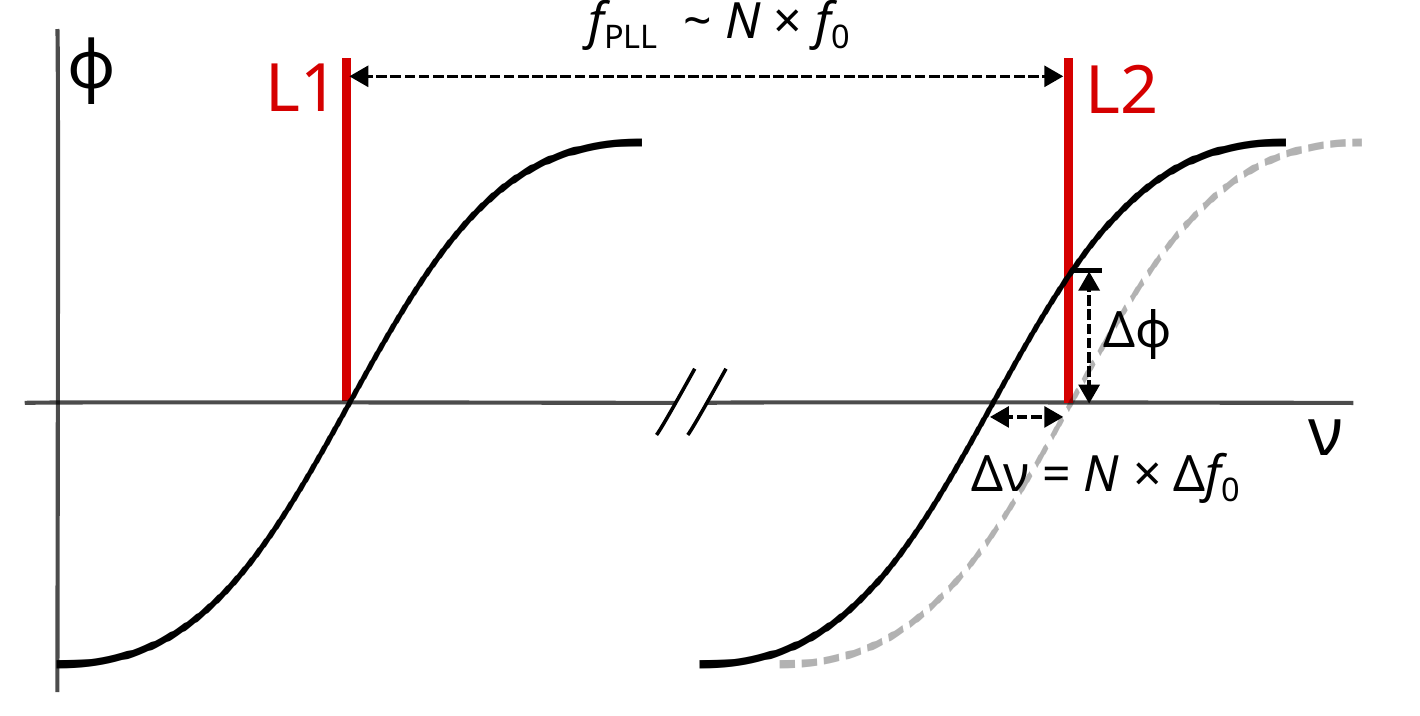}
    \caption{Conceptual overview of the measurement technique showing the relative phase in transmission of an optical cavity as a function of frequency near different resonance frequencies (black curves). Two laser fields are phase-locked such that they maintain a fixed relative frequency separation $f_{\rm PLL}$, initially selected such that the second laser is very near to another cavity resonance frequency (dashed curve). As the cavity length changes, $f_0$ changes and the resonance frequency with respect to the second laser is shifted by $\Delta \nu = N \times \Delta f_0$. The now-detuned second laser acquires an additional phase $\Delta \phi$ approximately proportional to the cavity length change.}
    \label{fig:theoreticalconcept}
\end{figure}

To measure the changing value of $f_0$ caused by variation in the cavity length, two laser fields are used: one which is actively stabilized to resonance, and another which is offset from the first by a fixed frequency such that it is approximately on resonance an integer number $N$ of FSRs away. By measuring the relative phase between these two laser fields in transmission of a cavity and knowing the initial value of the free spectral range, a timeseries of the absolute length of the cavity can be measured. The measurement concept is illustrated in Figure \ref{fig:theoreticalconcept}. 

With the field within the cavity resonance linewidth $\delta \nu \equiv f_0 / \mathcal{F}$, where $\Delta \nu \ll \delta \nu$, the phase component of the complex transmission coefficient can be expressed as,
\begin{align}
    \Delta \phi \approx 2 \mathcal{F} \frac{\Delta\nu}{f_0} = 2\mathcal{F} N \frac{\Delta f_0}{f_0}
\end{align}
where $\Delta f_0$ is the shift in the cavity free spectral range (such that, at a separation of $N$ FSRs, two resonance frequencies will shift by $\Delta\nu = N\Delta f_0$). Using the relation $\Delta f_0 / f_0 = -\Delta L / L$, the strain can be expressed as
\begin{align}
    \frac{\Delta L}{L} = -\frac{\Delta \phi}{2 \mathcal{F} \cdot N}
\end{align}
The sensitivity of the measurement to small variations in strain is therefore large for high finesse, long-baseline optical cavities. The requirement for the measured field to stay within the cavity resonance sets a range limit for the technique, such that the cavity length change does not cause the resonance frequency $N$ FSRs away to shift by more than a linewidth. To ensure the linearity of the measurement and the validity of the calibration, we further limit this to 1/10th of a cavity linewidth where deviations from the linear approximation are below 1\%. Therefore,
\begin{equation}\label{eq:range}
    \Delta L ^{\rm max} = L \frac{\delta \nu / 10}{N f_0}
\end{equation}

An alternative but related method for measuring the strain of a cavity using two lasers is to separately frequency stabilize each field on resonance $N$ free spectral ranges apart. The relative frequency between these fields is continuously measured, revealing the cavity FSR, and therefore the absolute cavity length, over time. 

\section{Experimental Design}

\begin{figure}
    \centering
    \includegraphics[width=0.8\linewidth]{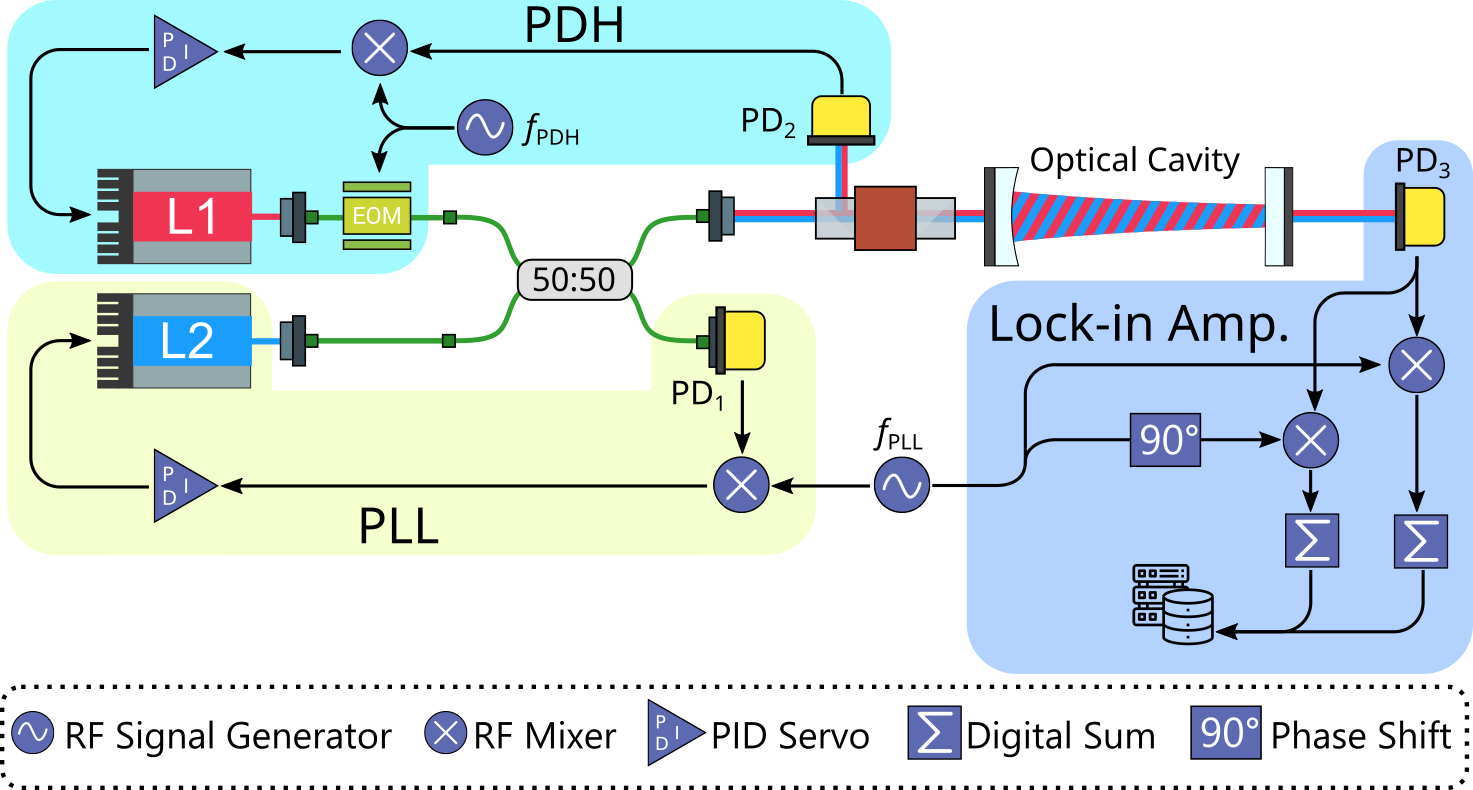}
    \caption{Layout of the cavity absolute length measurement concept. Elements of the PDH and PLL laser control loops are highlighted. The lock-in amplifier which records the heterodyne beat-note between the two lasers in transmission of the cavity is highlighted in blue.}
    \label{fig:experimentalconcept}
\end{figure}

The cavity absolute length measurement scheme is depicted in Figure \ref{fig:experimentalconcept}. Two continuous-wave lasers (Nd:YAG non-planar ring oscillator, 1064\,nm wavelength) located on the same side of the optical cavity are coupled into optical fibers. One laser (L1) is phase-modulated by a fiber electro-optic modulator (EOM) as part of a Pound-Drever-Hall (PDH) frequency stabilization loop \cite{drever1983laser, black2001introduction}. The two lasers are spatially combined in a fiber 50:50 beam-splitter, one port of which is directed towards a series of lenses and mirrors which mode-match and align the interfered beams to the cavity. A Faraday isolator directs light in reflection of the cavity onto a photodetector (PD2) which is digitally demodulated at the EOM driving frequency to form a PDH error signal which is used to control the frequency of L1. In this way, L1 is kept on resonance with the cavity. 

The second port of the fiber beam-splitter exits onto a photodetector (PD1), where the L1-L2 interference beat-note is sensed. This signal is mixed with a local oscillator at frequency $f_{\rm PLL}$ to form a phase-locked loop stabilizing the frequency and phase of L2 relative to L1. A photodetector on the opposite side of the cavity (PD3) measures the interference of the lasers in transmission and is measured by an FPGA-based lock-in amplifier (LIA). The LIA demodulates the transmission beat-note at the known relative laser frequency $f_{\rm PLL}$ (maintained by the PLL) in quadrature components. As the absolute length of the cavity changes, L2 detunes from resonance and acquires a phase shift which is measured by the lock-in amplifier and calibrated in terms of cavity length. The frequencies of all RF signals are synchronized to a common rubidium master oscillator which is referenced to GPS time, providing better than $10^{-14}$ timing accuracy.

\begin{figure}
    \centering
    \includegraphics[width=0.8\linewidth]{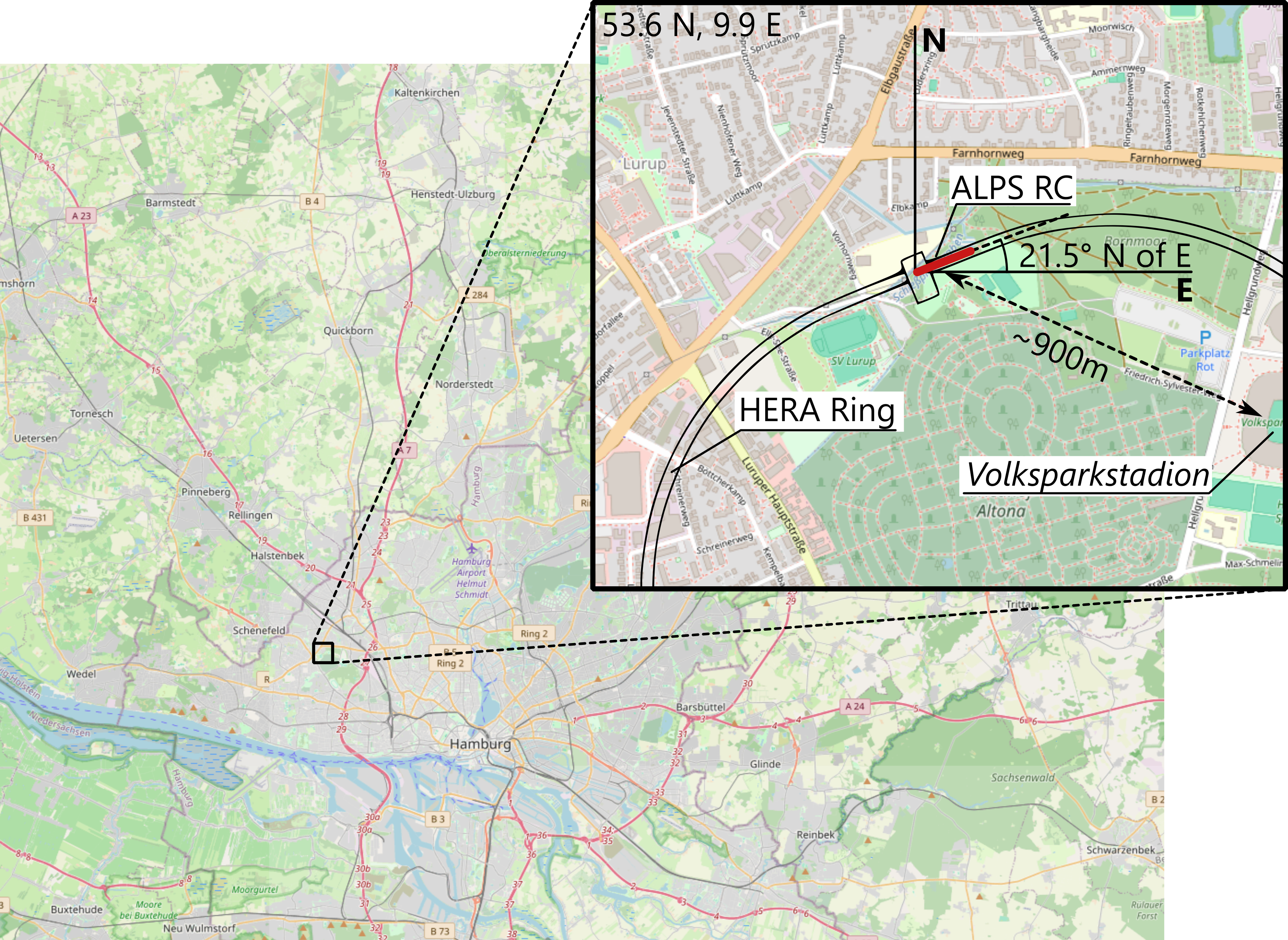}
    \caption{Map of the ALPS\,II regeneration cavity (in red) shown in the context of the HERA accelerator tunnel and immediate surrounds (inset) and the wider Hamburg area. The ALPS\,II experiment is oriented at 21.5$^\circ$ with respect to the east-west axis. Map data © OpenStreetMap contributors.}
    \label{fig:map}
\end{figure}

To demonstrate this technique, we used the ALPS\,II Regeneration Cavity (RC) \cite{Kozlowski2024} as our optical cavity. This 123-meter long, high-finesse optical cavity is located approximately 26\,m underground in a section of the tunnel formerly occupied by the HERA particle accelerator experiment in Hamburg, Germany (53.59 N, 9.88 E). The RC is aligned in a mostly east-west direction, with a compass angle of $21.5^\circ\pm1^\circ$ north from east. A map of the ALPS\,II RC in the context of the local environment is shown in Figure \ref{fig:map}. The mirrors that form the cavity are mounted to optical tables which sit on concrete blocks rigidly fixed to the tunnel. The mechanical transfer functions between the tunnel floor and each of the optical tables are measured to be unity below $\sim$\,5 Hz \cite{miller2020}, ensuring that the strain recorded is representative of the seismic strain in the accelerator tunnel.

The length of the RC can be adjusted by moving the optical breadboard to which the flat RC mirror is mounted using precision motors with 100\,nm resolution and out-of-loop position encoders. The size of the adjustment is further verified by measuring the cavity FSR before and after repositioning.

The finesse of the ALPS\,II cavity used for the following results was routinely measured via cavity ring-down measurements in order to accurately calibrate the recorded strain. The typical finesse was $\sim25,000\pm1,000$, measured before and after each of the following results. Similarly, the FSR spacing $N$, which also factors into the calibration, was selected depending on the measurement, based on a trade-off between sensitivity and the range over which the strain signal stays within the linear slope of the phase response. We selected $N = 80$ for the measurements in Sec. \ref{taylorswift} and $N = 50$ for the results in Sec. \ref{earthquake} and \ref{tides}. 

\section{Results and Discussion}

\subsection{Anthropogenic Noise from Taylor Swift Concert}
\label{taylorswift}
\begin{figure}
    \centering
    \includegraphics[width=0.9\linewidth]{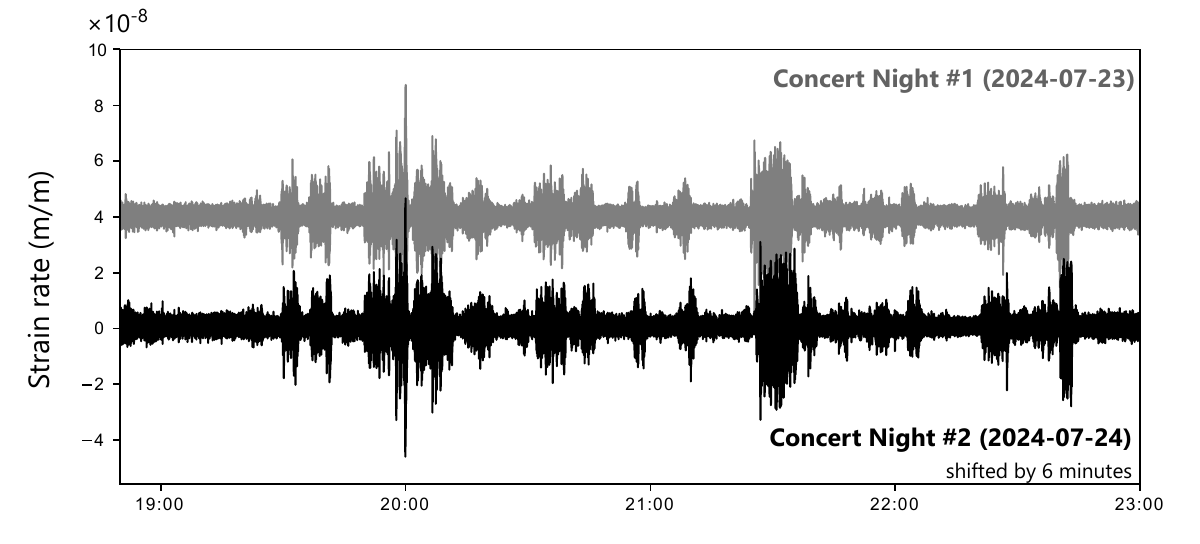}
    \caption{Measured strain of the ALPS\,II regeneration cavity on two evenings coinciding with the Taylor Swift concerts in the Volksparkstadion. The concerts nominally began at UTC19:30 each evening. The time axis of the second night is shifted forward by six minutes in order to emphasize the timing consistency between the two performances. An offset of $4\times10^{-8}$ strain is applied to the first night for visual clarity.}
    \label{fig:taylorswift_timeseries}
\end{figure}

\begin{figure}
    \centering
    \includegraphics[width=0.9\linewidth]{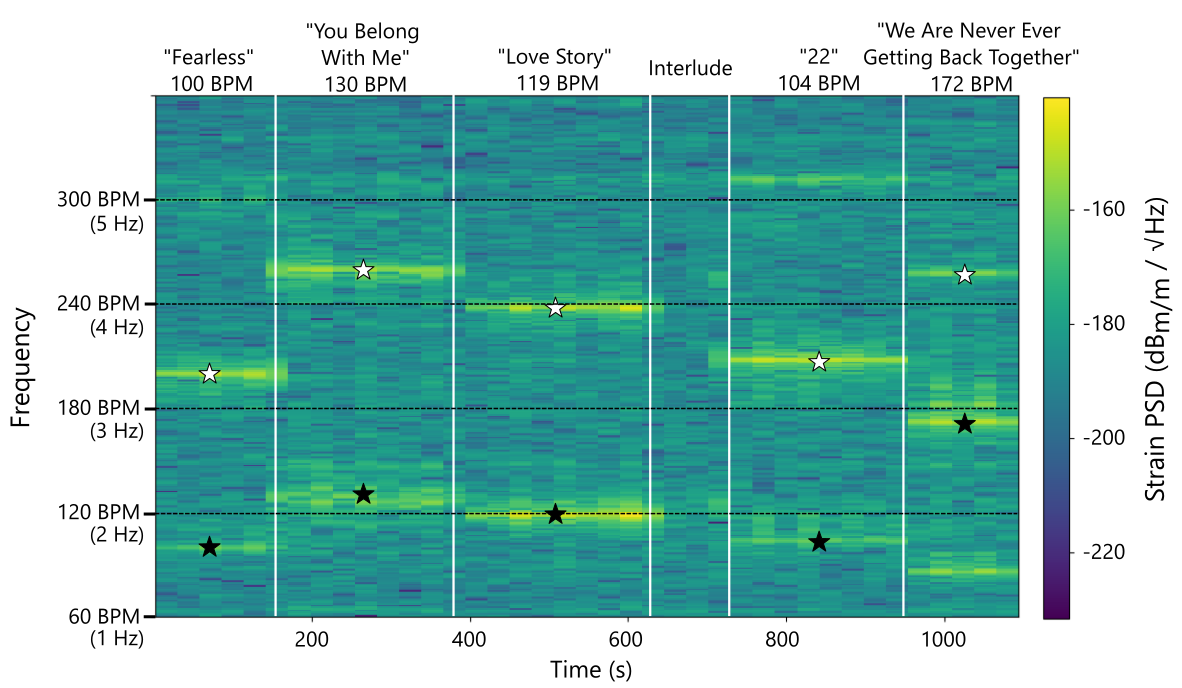}
    \caption{A representative twenty minute spectrogram sample of the cavity strain measured during the Taylor Swift concert. Five distinct songs which were performed during this period are denoted along with their published tempos (in beats-per-minute) \cite{songbpm}. Black stars are placed at the published tempo at each song, and white stars at their frequency-doubled values. These markers closely correspond to the frequencies with the highest measured strain power spectral densities. The third harmonic of "22" and the half-harmonic of "We Are Never Ever Getting Back Together" can also be seen prominently at 112\,BPM and 86\,BPM, respectively.}
    \label{fig:spectrogram}
\end{figure}

On the evenings of 23rd and 24th July 2024, Taylor Swift performed for 50,000 fans at the Hamburg Volksparkstadion as part of the "Eras Tour". Figure \ref{fig:taylorswift_timeseries} shows the measured strain in the ALPS\,II Regeneration Cavity coinciding with the concert, located approximately 1.1\,km from the midpoint of the cavity. Previous studies have recorded and analyzed the ground motion associated with the Taylor Swift concerts in Seattle, Washington \cite{Caplan2024}, Los Angeles, California \cite{tepp2024}, and Dublin \cite{Dunn2026}. Analysis of the measured strain agrees with many of the conclusions of these studies. We found that peaks in the spectral density (as seen in the spectrogram in Figure \ref{fig:spectrogram}) match the published beats-per-minute (BPM) tempos of the individual songs. This indicates coordinated crowd motion at the fundamental and/or multiple harmonic of the song's BPM.

The performances from each night are nearly identical in their seismic signatures - indeed, comparing individual songs from one night to the next yields correlation coefficients as high as 0.96 over the entire duration of the song. This suggests that on these two separate nights, these two (at least partially non-overlapping) groups behaved in an extremely similar fashion during the concert. The consistency of timing of each song-related seismic "feature" across the three-hour performance, aligned often to within a minute, also indicates a well-rehearsed schedule. The highest amplitude strain was recorded during the track "Love Story" each night at approximately 20:00, whereas the track with the largest seismic energy integrated over its duration on both nights was "Shake It Off", beginning at approximately 21:25. The measurements of the crowd-induced ground motion match observations performed by the WAVE seismic distributed sensor network located at various locations at the Hamburg Bahrenfeld campus \cite{wave}.

\subsection{Earthquake}
\label{earthquake}
\begin{figure}
    \centering
    \includegraphics[width=0.9\linewidth]{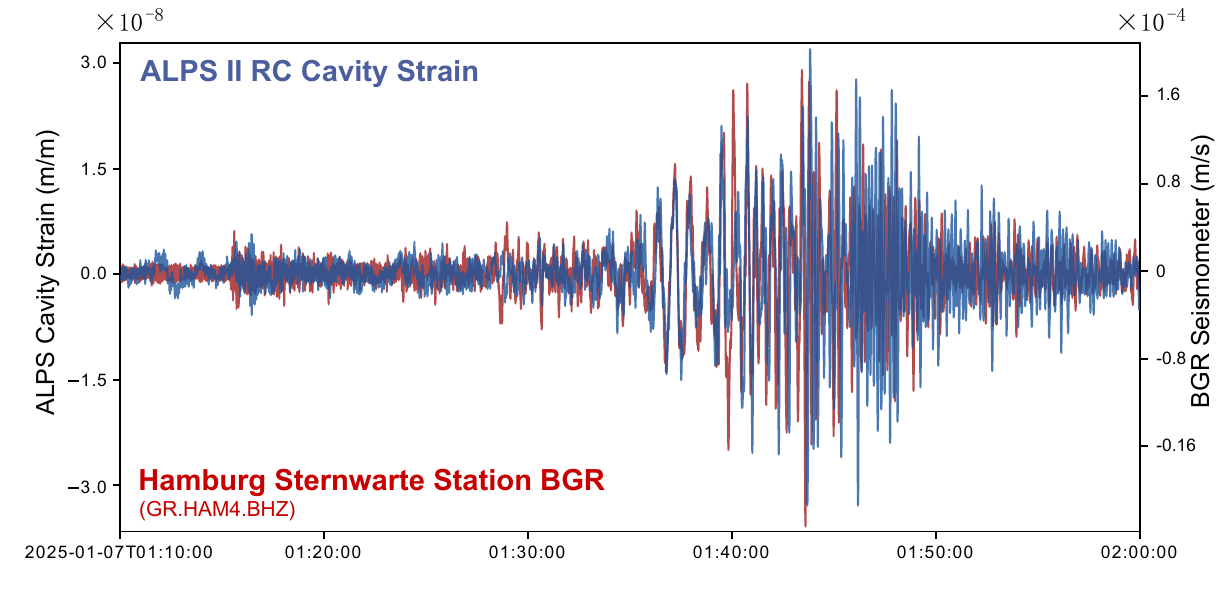}
    \caption{Measurement of the January 2025 Tibet Earthquake derived from the strain of the ALPS\,II regeneration cavity (blue) and recorded at the BGR Seismology Station Hamburger Sternwarte (red). In the first few minutes after the arrival of surface waves (UTC 01:35:00), the two re-scaled measurements are nearly  identical in behavior, with a cross-correlation coefficient above 0.90 with no time shift. Sternwarte data was obtained via the publicly accessible GRSN database \cite{grsn}.}
    \label{fig:earthquake}
\end{figure}

On 7th January 2025, during a long-term strain measurement, a transient event was recorded and later identified as the magnitude 7.1 Southern Tibetan Plateau Earthquake (UTC 2025-01-07 01:05:16, 28.573°N 87.375°E) \cite{tibet2025}. This deadly earthquake was recorded by seismology stations around the world, including at the nearby BGR Station Hamburger Sternwarte \cite{grsn}. The recorded RC strain and the seismometer data are shown in Figure \ref{fig:earthquake}. The two recordings are very similar, with the seismic waves arriving simultaneously (< 0.1\,s) and correlation coefficients exceeding 0.90 over multiple segments.

The arrival time of the dominant surface Rayleigh waves at approximately UTC 01:35:00 is consistent with the epicentral distance ($\sim$\,6,700\,km, $\Delta \approx60^\circ$) and the group velocity of these surface waves ($\sim$\,3.75\,km/s). Although much lower in strain amplitude, the arrival of P-waves at approximately UTC 01:14 can also be observed, once again consistent with propagation models and the results of local geo-seismic stations \cite{seismology}.

\subsection{Earth Tide}\label{tides}
\begin{figure}
    \centering
    \includegraphics[width=0.9\linewidth]{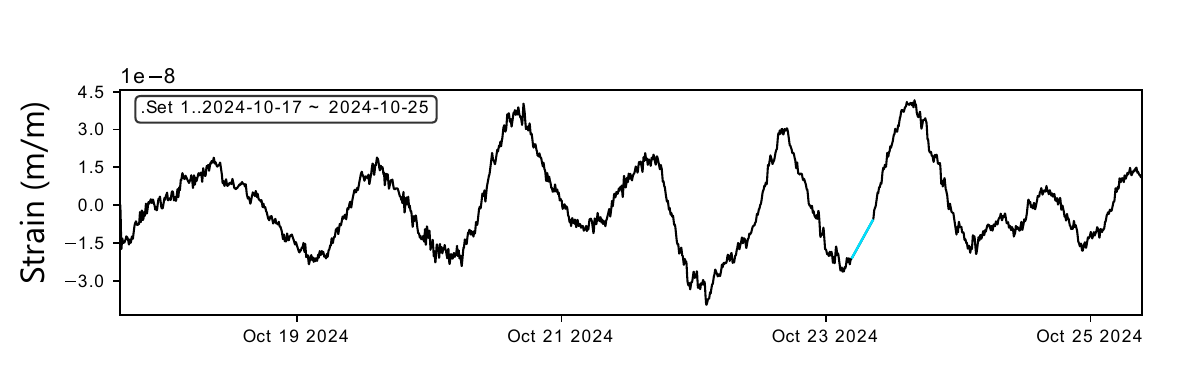}
    \caption{Regeneration cavity strain measured over a six day period in October. The strain data is linearly detrended and low-pass filtered at 10\,mHz. The straight segment on 23 October represents a linear interpolation applied to a period of extended system unlock. Oscillations are dominated by approximately daily variations correlated strongly with environmental conditions in the experimental hall, particularly temperature fluctuations.}
    \label{fig:longterm_timeseries}
\end{figure}

Understanding the behavior of the RC absolute length over multi-day periods is essential to the successful operation of the ALPS\,II experiment \cite{diazortiz2022}. The long-term length of the RC was measured in three one-week segments - beginning 17th October 2024, 2nd January 2025, and 17th January 2025 - for a total of 21 days and 14 hours of data. Utilizing an automated re-lock software interfaced with the PDH and PLL control loops, an average duty cycle of 94.3\% was achieved over the 21 days of measurement. Periods in which either the PLL or PDH are unlocked, typically less than 5 minutes, are automatically identified and replaced with linearly interpolated values. As discussed in Section \ref{sec:theory}, the linear scaling of phase to length is only valid when L1 remains within the cavity linewidth, limiting the range to (recalling Eq. \ref{eq:range}),
\begin{equation}
    \Delta L ^{\rm max} \approx 123\,\mathrm{m} \, \frac{4.8\,\mathrm{Hz}}{50 \times 1.22\,\mathrm{MHz}} = 9.7 \mathrm{\mu m}
\end{equation}
Very slow drifts of the cavity length over multiple days, typically due to local environmental effects, occasionally caused the cavity length drift to reach this range limitation. At each of these instances, the length of the RC is automatically adjusted using a motorized stage so that L1 is re-centered on resonance and the length correspondingly compensated, yielding a continuous time-series. The strain of the RC during the measurement period is shown in Figure \ref{fig:longterm_timeseries}, detrended and low-pass filtered at 0.1\,Hz.

\begin{figure}
    \centering
    \includegraphics[width=0.8\linewidth]{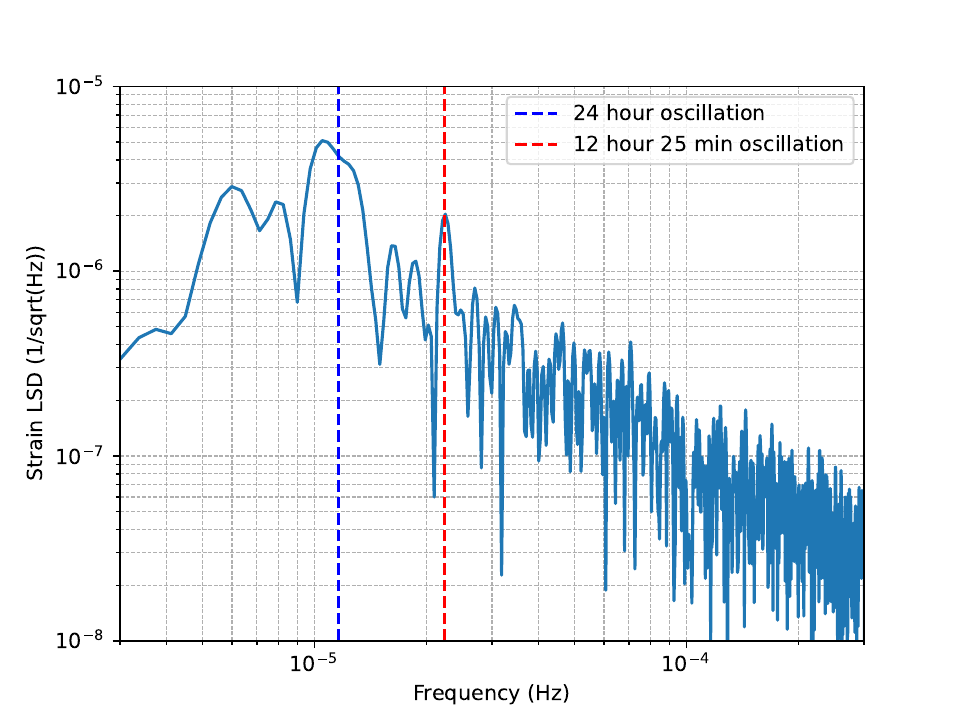}
    \caption{Linear spectral density of the combined 21 days of long-term cavity strain measurement. A 1/$f$ trend is present across the frequency spectrum, with a reduction below $6\,\mu\rm Hz$ resulting from a third-order polynomial de-trending of the time-series. Peaks associated with daily (24 hr) and semidiurnal lunar (12 hr 25 minutes) variations are identified.}
    \label{fig:tidalLSD}
\end{figure}

Fig. \ref{fig:tidalLSD} depicts the linear spectral density (LSD) of the RC strain over the full 21 days of measurement. We see a clear strain component at the frequency corresponding to the semidiurnal lunar earth tide (corresponding to the M2 and S2 tidal modes) and a peak at diurnal periods (corresponding primarily to the O1 and K1 modes)\cite{agnew2007}. No weaker tidal component was clearly discernible in our measurement. The measurement LSD is limited by experimental noise around the level of $10^{-9}$ strain at frequencies above 1\,mHz primarily due to the limited ability of the PDH frequency stabilization loop to stay perfectly on resonance, resulting in residual frequency noise on L1. Below 1\,mHz, this noise is substantially below the length noise of the cavity.

The amplitude of the semidiurnal peak is approximately $2\times10^{-9}$ strain rms. This small amplitude (relative to the semidiurnal earth tide measured by other laser strainmeter observations \cite{Araya1999,Takamori2014}) can be attributed to the experiment's location and compass orientation: in the simple model of an ocean-less earth, semidiurnal strain tides achieve a minimum value of zero at a latitude of 52.4$^\circ$ \cite{agnew2007}. This zero-crossing arises due to the $\cos^2{\phi} - 1/3$ latitude dependence of the horizontal strain. The ALPS\,II Regeneration Cavity is located at 53.6$^\circ$ N and oriented principally in the east-west direction. 

The theoretical tidal amplitudes for the location and orientation of the ALPS\,II experiment are modeled using \texttt{pygtide}, a python wrapper for the ETERNA-x earth tide simulation software \cite{ETERNA}, with additional effects from ocean-loading using "Some Programs for Ocean-Tide Loading (SPOTL)" \cite{SPOTL}. Body tide and ocean loading contributions are computed separately and summed per tidal constituent before projection onto the cavity orientation and location. The measured values of the earth tides from the RC strainmeter measurement are compared with these models, with the results shown in Table \ref{tab:comparison-to-model}. The rms values of the model and the measurements were calculated by integrating the LSD in each component band, using the same bandwidth for the integration in both model and experimental data.

The modeling involves several simplifying assumptions, including the choice of Earth model Green's functions, and does not account for atmospheric loading or direct thermal effects on the cavity; given these factors, the present agreement is considered satisfactory. A more thorough consideration of the deviations from the model is outside the scope of this paper.

\begin{table}[]
\centering
\begin{tabular}{ |p{3.5cm}||p{3cm}|p{3cm}|p{1.5cm}|  }
 \hline
 Period Band & Model strain (rms) & Measured strain (rms) & Deviation\\
 \hline
 Diurnal ($\sim24$ hrs)  & $5.80\times10^{-9}$   & $7.50(\pm2.97) \times10^{-9}$ & +29\% \\
 Semidiurnal ($\sim12$ hrs) &   $1.76\times10^{-9}$  & $2.06(\pm0.67)\times10^{-9}$  & +17\% \\
 \hline
\end{tabular}
    \caption{Comparison of modeled and measured tidal strain amplitudes in the two primary bands. The error represents a $1\sigma$ uncertainty derived from the noise in adjacent spectral bins, as well as a 5\% systematic uncertainty in the calibration.}
    \label{tab:comparison-to-model}
\end{table}

\section{Conclusions}
By measuring the relative phase change between two laser fields interacting with the 123-meter long optical cavity, we have turned a portion of the ALPS\,II particle search experiment into an accurate and precise strainmeter. Using this technique, we have observed a diverse set of transient seismic signals, including earthquakes and anthropogenic ground noise resulting from cultural events in Hamburg. We also recorded the ultra-low frequency strain of the earth due to tidal effects, despite the cavity's orientation and latitude placing it within $1.2^\circ$ of the semidiurnal tidal strain node, where the expected horizontal strain amplitude passes through zero. The alternative approach for measuring the changing cavity FSR using two fields individually stabilized to the cavity discussed in Section \ref{sec:theory} was demonstrated to have roughly equivalent sensitivity to length changes in a separate Fabry-Perot cavity experiment \cite{spector2025demonstrationinterferometrictechniquemeasuring}. This particular technique was not employed in our study as the ALPS\,II experiment is not equipped to simultaneously stabilize two lasers to the cavity.

The ALPS\,II RC was never designed to operate as a dedicated seismic strainmeter, and several aspects of the experiment could be improved to further optimize such measurements. A longer interferometer - for example, utilizing the full 250 meters of the ALPS\,II infrastructure - would improve sensitivity by increasing the strain baseline. Likewise, upgraded mirrors with higher reflectivity and lower losses are planned for ALPS\,II, which would further increase sensitivity. Finally, more robust climate monitoring and control along the length of the interferometer would contribute to reducing temperature-dependent cavity length changes. The future installation of distributed acoustic sensing fibers or seismometers co-located in the HERA North hall would contribute to better understanding of our sensitivity and provide a comparison for future measurements, as currently we are unable to distinguish if our sensitivity is limited by the experiment or the ambient seismic conditions of the hall.

\begin{backmatter}
\bmsection{Funding} We gratefully acknowledge the support of the National Science Foundation (Grant no. PHY-2309918), the Heising-Simons Foundation (Grant no. 2020-1841), the Deutsche Forschungsgemeinschaft under Germany’s Excellence Strategy (EXC 2121 “Quantum Universe" 390833306 and WI 1643/2-1), the Partnership for Innovation, Education and Research (PIER) of DESY and Universität Hamburg under PIER Seed Project (PIF-2022-18), the German Volkswagen Stiftung, and the UK Science and Technologies Facilities Council (Grant no. ST/T006331/1).
\bmsection{Acknowledgments} We are thankful to all members of the ALPS\,II collaboration, as well as to the site and engineering support from the Deutsches Elektronen-Synchrotron DESY. The authors would also like to thank Celine Hadziioannou, Katharina-Sophie Isleif, and the other members of the WAVE collaboration for their valuable insight.
\bmsection{Disclosures} The authors declare no conflicts of interest.
\bmsection{Data availability} Data underlying the results presented in this paper may be obtained from the authors upon reasonable request.

\end{backmatter}

\bibliography{absolute_length}

@article{Abbott2009,
doi = {10.1088/0034-4885/72/7/076901},
url = {https://dx.doi.org/10.1088/0034-4885/72/7/076901},
year = {2009},
month = {jun},
publisher = {},
volume = {72},
number = {7},
pages = {076901},
author = {B P Abbott and R Abbott and R Adhikari and P Ajith and B Allen and G Allen and R S Amin and S B Anderson and W G Anderson and M A Arain and M Araya and H Armandula and P Armor and Y Aso and S Aston and P Aufmuth and C Aulbert and S Babak and P Baker and S Ballmer and C Barker and D Barker and B Barr and P Barriga and L Barsotti and M A Barton and I Bartos and R Bassiri and M Bastarrika and B Behnke and M Benacquista and J Betzwieser and P T Beyersdorf and I A Bilenko and G Billingsley and R Biswas and E Black and J K Blackburn and L Blackburn and D Blair and B Bland and T P Bodiya and L Bogue and R Bork and V Boschi and S Bose and P R Brady and V B Braginsky and J E Brau and D O Bridges and M Brinkmann and A F Brooks and D A Brown and A Brummit and G Brunet and A Bullington and A Buonanno and O Burmeister and R L Byer and L Cadonati and J B Camp and J Cannizzo and K C Cannon and J Cao and L Cardenas and S Caride and G Castaldi and S Caudill and M Cavaglià and C Cepeda and T Chalermsongsak and E Chalkley and P Charlton and S Chatterji and S Chelkowski and Y Chen and N Christensen and C T Y Chung and D Clark and J Clark and J H Clayton and T Cokelaer and C N Colacino and R Conte and D Cook and T R C Corbitt and N Cornish and D Coward and D C Coyne and J D E Creighton and T D Creighton and A M Cruise and R M Culter and A Cumming and L Cunningham and S L Danilishin and K Danzmann and B Daudert and G Davies and E J Daw and D DeBra and J Degallaix and V Dergachev and S Desai and R DeSalvo and S Dhurandhar and M Díaz and A Dietz and F Donovan and K L Dooley and E E Doomes and R W P Drever and J Dueck and I Duke and J-C Dumas and J G Dwyer and C Echols and M Edgar and A Effler and P Ehrens and E Espinoza and T Etzel and M Evans and T Evans and S Fairhurst and Y Faltas and Y Fan and D Fazi and H Fehrmenn and L S Finn and K Flasch and S Foley and C Forrest and N Fotopoulos and A Franzen and M Frede and M Frei and Z Frei and A Freise and R Frey and T Fricke and P Fritschel and V V Frolov and M Fyffe and V Galdi and J A Garofoli and I Gholami and J A Giaime and S Giampanis and K D Giardina and K Goda and E Goetz and L M Goggin and G González and M L Gorodetsky and S Goßler and R Gouaty and A Grant and S Gras and C Gray and M Gray and R J S Greenhalgh and A M Gretarsson and F Grimaldi and R Grosso and H Grote and S Grunewald and M Guenther and E K Gustafson and R Gustafson and B Hage and J M Hallam and D Hammer and G D Hammond and C Hanna and J Hanson and J Harms and G M Harry and I W Harry and E D Harstad and K Haughian and K Hayama and J Heefner and I S Heng and A Heptonstall and M Hewitson and S Hild and E Hirose and D Hoak and K A Hodge and K Holt and D J Hosken and J Hough and D Hoyland and B Hughey and S H Huttner and D R Ingram and T Isogai and M Ito and A Ivanov and B Johnson and W W Johnson and D I Jones and G Jones and R Jones and L Ju and P Kalmus and V Kalogera and S Kandhasamy and J Kanner and D Kasprzyk and E Katsavounidis and K Kawabe and S Kawamura and F Kawazoe and W Kells and D G Keppel and A Khalaidovski and F Y Khalili and R Khan and E Khazanov and P King and J S Kissel and S Klimenko and K Kokeyama and V Kondrashov and R Kopparapu and S Koranda and D Kozak and B Krishnan and R Kumar and P Kwee and P K Lam and M Landry and B Lantz and A Lazzarini and H Lei and M Lei and N Leindecker and I Leonor and C Li and H Lin and P E Lindquist and T B Littenberg and N A Lockerbie and D Lodhia and M Longo and M Lormand and P Lu and M Lubinski and A Lucianetti and H Lück and B Machenschalk and M MacInnis and M Mageswaran and K Mailand and I Mandel and V Mandic and S Márka and Z Márka and A Markosyan and J Markowitz and E Maros and I W Martin and R M Martin and J N Marx and K Mason and F Matichard and L Matone and R A Matzner and N Mavalvala and R McCarthy and D E McClelland and S C McGuire and M McHugh and G McIntyre and D J A McKechan and K McKenzie and M Mehmet and A Melatos and A C Melissinos and D F Menéndez and G Mendell and R A Mercer and S Meshkov and C Messenger and M S Meyer and J Miller and J Minelli and Y Mino and V P Mitrofanov and G Mitselmakher and R Mittleman and O Miyakawa and B Moe and S D Mohanty and S R P Mohapatra and G Moreno and T Morioka and K Mors and K Mossavi and C MowLowry and G Mueller and H Müller-Ebhardt and D Muhammad and S Mukherjee and H Mukhopadhyay and A Mullavey and J Munch and P G Murray and E Myers and J Myers and T Nash and J Nelson and G Newton and A Nishizawa and K Numata and J O'Dell and B O'Reilly and R O'Shaughnessy and E Ochsner and G H Ogin and D J Ottaway and R S Ottens and H Overmier and B J Owen and Y Pan and C Pankow and M A Papa and V Parameshwaraiah and P Patel and M Pedraza and S Penn and A Perraca and V Pierro and I M Pinto and M Pitkin and H J Pletsch and M V Plissi and F Postiglione and M Principe and R Prix and L Prokhorov and O Punken and V Quetschke and F J Raab and D S Rabeling and H Radkins and P Raffai and Z Raics and N Rainer and M Rakhmanov and V Raymond and C M Reed and T Reed and H Rehbein and S Reid and D H Reitze and R Riesen and K Riles and B Rivera and P Roberts and N A Robertson and C Robinson and E L Robinson and S Roddy and C Röver and J Rollins and J D Romano and J H Romie and S Rowan and A Rüdiger and P Russell and K Ryan and S Sakata and L Sancho de la Jordana and V Sandberg and V Sannibale and L Santamaría and S Saraf and P Sarin and B S Sathyaprakash and S Sato and M Satterthwaite and P R Saulson and R Savage and P Savov and M Scanlan and R Schilling and R Schnabel and R Schofield and B Schulz and B F Schutz and P Schwinberg and J Scott and S M Scott and A C Searle and B Sears and F Seifert and D Sellers and A S Sengupta and A Sergeev and B Shapiro and P Shawhan and D H Shoemaker and A Sibley and X Siemens and D Sigg and S Sinha and A M Sintes and B J J Slagmolen and J Slutsky and J R Smith and M R Smith and N D Smith and K Somiya and B Sorazu and A Stein and L C Stein and S Steplewski and A Stochino and R Stone and K A Strain and S Strigin and A Stroeer and A L Stuver and T Z Summerscales and K-X Sun and M Sung and P J Sutton and G P Szokoly and D Talukder and L Tang and D B Tanner and S P Tarabrin and J R Taylor and R Taylor and J Thacker and K A Thorne and A Thüring and K V Tokmakov and C Torres and C Torrie and G Traylor and M Trias and D Ugolini and J Ulmen and K Urbanek and H Vahlbruch and M Vallisneri and C Van Den Broeck and M V van der Sluys and A A van Veggel and S Vass and R Vaulin and A Vecchio and J Veitch and P Veitch and C Veltkamp and A Villar and C Vorvick and S P Vyachanin and S J Waldman and L Wallace and R L Ward and A Weidner and M Weinert and A J Weinstein and R Weiss and L Wen and S Wen and K Wette and J T Whelan and S E Whitcomb and B F Whiting and C Wilkinson and P A Willems and H R Williams and L Williams and B Willke and I Wilmut and L Winkelmann and W Winkler and C C Wipf and A G Wiseman and G Woan and R Wooley and J Worden and W Wu and I Yakushin and H Yamamoto and Z Yan and S Yoshida and M Zanolin and J Zhang and L Zhang and C Zhao and N Zotov and M E Zucker and H zur Mühlen and J Zweizig},
title = {LIGO: the Laser Interferometer Gravitational-Wave Observatory},
journal = {Reports on Progress in Physics}
}

@article{Accadia2012,
doi = {10.1088/1748-0221/7/03/P03012},
url = {https://dx.doi.org/10.1088/1748-0221/7/03/P03012},
year = {2012},
month = {mar},
publisher = {},
volume = {7},
number = {03},
pages = {P03012},
author = {T Accadia and  F Acernese and  M Alshourbagy and  P Amico and  F Antonucci and  S Aoudia and  N Arnaud and  C Arnault and  K G Arun and  P Astone and  S Avino and  D Babusci and  G Ballardin and  F Barone and  G Barrand and  L Barsotti and  M Barsuglia and  A Basti and  Th S Bauer and  F Beauville and  M Bebronne and  M Bejger and  M G Beker and  F Bellachia and  A Belletoile and  J L Beney and  M Bernardini and  S Bigotta and  R Bilhaut and  S Birindelli and  M Bitossi and  M A Bizouard and  M Blom and  C Boccara and  D Boget and  F Bondu and  L Bonelli and  R Bonnand and  V Boschi and  L Bosi and  T Bouedo and  B Bouhou and  A Bozzi and  L Bracci and  S Braccini and  C Bradaschia and  M Branchesi and  T Briant and  A Brillet and  V Brisson and  L Brocco and  T Bulik and  H J Bulten and  D Buskulic and  C Buy and  G Cagnoli and  G Calamai and  E Calloni and  E Campagna and  B Canuel and  F Carbognani and  L Carbone and  F Cavalier and  R Cavalieri and  R Cecchi and  G Cella and  E Cesarini and  E Chassande-Mottin and  S Chatterji and  R Chiche and  A Chincarini and  A Chiummo and  N Christensen and  A C Clapson and  F Cleva and  E Coccia and  P -F Cohadon and  C N Colacino and  J Colas and  A Colla and  M Colombini and  G Conforto and  A Corsi and  S Cortese and  F Cottone and  J -P Coulon and  E Cuoco and  S D'Antonio and  G Daguin and  A Dari and  V Dattilo and  P Y David and  M Davier and  R Day and  G Debreczeni and  G De Carolis and  M Dehamme and  R Del Fabbro and  W Del Pozzo and  M del Prete and  L Derome and  R De Rosa and  R DeSalvo and  M Dialinas and  L Di Fiore and  A Di Lieto and  M Di Paolo Emilio and  A Di Virgilio and  A Dietz and  M Doets and  P Dominici and  A Dominjon and  M Drago and  C Drezen and  B Dujardin and  B Dulach and  C Eder and  A Eleuteri and  D Enard and  M Evans and  L Fabbroni and  V Fafone and  H Fang and  I Ferrante and  F Fidecaro and  I Fiori and  R Flaminio and  D Forest and  L A Forte and  J -D Fournier and  L Fournier and  J Franc and  O Francois and  S Frasca and  F Frasconi and  A Freise and  A Gaddi and  M Galimberti and  L Gammaitoni and  P Ganau and  C Garnier and  F Garufi and  M E Gáspár and  G Gemme and  E Genin and  A Gennai and  G Gennaro and  L Giacobone and  A Giazotto and  G Giordano and  L Giordano and  C Girard and  R Gouaty and  A Grado and  M Granata and  V Granata and  X Grave and  C Greverie and  H Groenstege and  G M Guidi and  S Hamdani and  J -F Hayau and  S Hebri and  A Heidmann and  H Heitmann and  P Hello and  G Hemming and  E Hennes and  R Hermel and  P Heusse and  L Holloway and  D Huet and  M Iannarelli and  P Jaranowski and  D Jehanno and  L Journet and  S Karkar and  T Ketel and  H Voet and  J Kovalik and  I Kowalska and  S Kreckelbergh and  A Krolak and  J C Lacotte and  B Lagrange and  P La Penna and  M Laval and  J C Le Marec and  N Leroy and  N Letendre and  T G F Li and  B Lieunard and  N Liguori and  O Lodygensky and  B Lopez and  M Lorenzini and  V Loriette and  G Losurdo and  M Loupias and  J M Mackowski and  T Maiani and  E Majorana and  C Magazzù and  I Maksimovic and  V Malvezzi and  N Man and  S Mancini and  B Mansoux and  M Mantovani and  F Marchesoni and  F Marion and  P Marin and  J Marque and  F Martelli and  A Masserot and  L Massonnet and  G Matone and  L Matone and  M Mazzoni and  F Menzinger and  C Michel and  L Milano and  Y Minenkov and  S Mitra and  M Mohan and  J -L Montorio and  R Morand and  F Moreau and  J Moreau and  N Morgado and  A Morgia and  S Mosca and  V Moscatelli and  B Mours and  P Mugnier and  F -A Mul and  L Naticchioni and  I Neri and  F Nocera and  E Pacaud and  G Pagliaroli and  A Pai and  L Palladino and  C Palomba and  F Paoletti and  R Paoletti and  A Paoli and  S Pardi and  G Parguez and  M Parisi and  A Pasqualetti and  R Passaquieti and  D Passuello and  M Perciballi and  B Perniola and  G Persichetti and  S Petit and  M Pichot and  F Piergiovanni and  M Pietka and  R Pignard and  L Pinard and  R Poggiani and  P Popolizio and  T Pradier and  M Prato and  G A Prodi and  M Punturo and  P Puppo and  K Qipiani and  O Rabaste and  D S Rabeling and  I Rácz and  F Raffaelli and  P Rapagnani and  S Rapisarda and  V Re and  A Reboux and  T Regimbau and  V Reita and  A Remilleux and  F Ricci and  I Ricciardi and  F Richard and  M Ripepe and  F Robinet and  A Rocchi and  L Rolland and  R Romano and  D Rosińska and  P Roudier and  P Ruggi and  G Russo and  L Salconi and  V Sannibale and  B Sassolas and  D Sentenac and  S Solimeno and  R Sottile and  L Sperandio and  R Stanga and  R Sturani and  B Swinkels and  M Tacca and  R Taddei and  L Taffarello and  M Tarallo and  S Tissot and  A Toncelli and  M Tonelli and  O Torre and  E Tournefier and  F Travasso and  C Tremola and  E Turri and  G Vajente and  J F J van den Brand and  C Van Den Broeck and  S van der Putten and  M Vasuth and  M Vavoulidis and  G Vedovato and  D Verkindt and  F Vetrano and  O Véziant and  A Viceré and  J -Y Vinet and  S Vilalte and  S Vitale and  H Vocca and  R L Ward and  M Was and  K Yamamoto and  M Yvert and  J -P Zendri and  Z Zhang},
title = {Virgo: a laser interferometer to detect gravitational waves},
journal = {Journal of Instrumentation}
}

@Article{Takamori2014,
AUTHOR = {Takamori, Akiteru and Araya, Akito and Morii, Wataru and Telada, Souichi and Uchiyama, Takashi and Ohashi, Masatake},
TITLE = {A 100-m Fabry–Pérot Cavity with Automatic Alignment Controls for Long-Term Observations of Earth’s Strain},
JOURNAL = {Technologies},
VOLUME = {2},
YEAR = {2014},
NUMBER = {3},
PAGES = {129--142},
URL = {https://www.mdpi.com/2227-7080/2/3/129},
ISSN = {2227-7080},
DOI = {10.3390/technologies2030129}
}

@article{Staley2015,
author = {A. Staley and D. Hoak and A. Effler and K. Izumi and S. Dwyer and K. Kawabe and E. J. King and M. Rakhmanov and R. L. Savage and D. Sigg},
journal = {Opt. Express},
number = {15},
pages = {19417--19431},
publisher = {Optica Publishing Group},
title = {High precision optical cavity length and width measurements using double modulation},
volume = {23},
month = {Jul},
year = {2015},
url = {https://opg.optica.org/oe/abstract.cfm?URI=oe-23-15-19417},
doi = {10.1364/OE.23.019417}
}

@article{Araya1999,
author = {Akito Araya and Souichi Telada and Kuniharu Tochikubo and Shinsuke Taniguchi and Ryutaro Takahashi and Keita Kawabe and Daisuke Tatsumi and Toshitaka Yamazaki and Seiji Kawamura and Shinji Miyoki and Shigenori Moriwaki and Mitsuru Musha and Shigeo Nagano and Masa-Katsu Fujimoto and Kazuo Horikoshi and Norikatsu Mio and Yutaka Naito and Akiteru Takamori and Kazuhiro Yamamoto},
journal = {Appl. Opt.},
number = {13},
pages = {2848--2856},
publisher = {Optica Publishing Group},
title = {Absolute-length determination of a long-baseline Fabry--Perot cavity by means ofresonating modulation sidebands},
volume = {38},
month = {May},
year = {1999},
url = {https://opg.optica.org/ao/abstract.cfm?URI=ao-38-13-2848},
doi = {10.1364/AO.38.002848},
}

@article{Caplan2024,
    author = {Jacqueline Caplan-Auerbach and Kyla Marczewski and Gavin S. Bullock},
    title = {Beast Quake (Taylor’s Version): Analysis
of Seismic Signals Recorded during Two
Taylor Swift Concerts in Seattle, July 2023},
    journal = {GSA Today},
    year = {2024},
    month = {May},
    volume = {34},
    issue = {5},
    pages = {4-10},
    doi = {10.1130/GSATG589A.1}
}

@article{diazortiz2022,
title = {Design of the ALPS II optical system},
journal = {Physics of the Dark Universe},
volume = {35},
pages = {100968},
year = {2022},
issn = {2212-6864},
doi = {https://doi.org/10.1016/j.dark.2022.100968},
url = {https://www.sciencedirect.com/science/article/pii/S2212686422000115},
author = {M. {Diaz Ortiz} and J. Gleason and H. Grote and A. Hallal and M.T. Hartman and H. Hollis and K.-S. Isleif and A. James and K. Karan and T. Kozlowski and A. Lindner and G. Messineo and G. Mueller and J.H. Põld and R.C.G. Smith and A.D. Spector and D.B. Tanner and L.-W. Wei and B. Willke}
}

@misc{Kozlowski2024,
      title={Design and Performance of the ALPS II Regeneration Cavity}, 
      author={Todd Kozlowski and Li-Wei Wei and Aaron D. Spector and Ayman Hallal and Henry Fraedrich and Daniel C. Brotherton and Isabella Oceano and Aldo Ejlli and Hartmut Grote and Harold Hollis and Kanioar Karan and Guido Mueller and D. B. Tanner and Benno Willke and Axel Lindner},
      year={2024},
      eprint={2408.13218},
      archivePrefix={arXiv},
      primaryClass={physics.optics},
      url={https://arxiv.org/abs/2408.13218}, 
}

@INCOLLECTION{agnew2007,
       author = {{Agnew}, D.~C.},
        title = "{Earth Tides}",
    booktitle = {Geodesy},
         year = 2007,
       editor = {{Schubert}, Gerald},
       volume = {3},
        pages = {163-195},
          doi = {10.1016/B978-044452748-6.00056-0},
       adsurl = {https://ui.adsabs.harvard.edu/abs/2007gdsy.book..163A}
}

@phdthesis{miller2020,
    author = {D. Miller},
    title = {Seismic noise analysis and isolation concepts for the ALPS II experiment at DESY},
    school = {Leibniz Universität Hannover},
    year = {2020}
}

@misc{songbpm,
  title = {SongBPM},
  howpublished = {\url{http://songbpm.com}},
  note = {Accessed: 2025-02-17}
}

@misc{wave,
  title = {WAVE initiative: First results from the Taylor Swift concert!},
  howpublished = {\url{https://wave-hamburg.eu/results/TaylorSwiftResults1/}},
  note = {Accessed: 2025-05-17}
}

@misc{grsn,
    title = {Federal Institute for Geosciences and Natural Resources: German Regional Seismic Network (GRSN)},
    howpublished = {doi: 10 .25928/mbx6-hr74},
    note = {Accessed: 2025-02-18}
}

@article{TAKEMOTO2004,
title = {A 100m laser strainmeter system installed in a 1km deep tunnel at Kamioka, Gifu, Japan},
journal = {Journal of Geodynamics},
volume = {38},
number = {3},
pages = {477-488},
year = {2004},
note = {Time Varying Gravimenty, GGP, and Vertical Crustal Motions},
issn = {0264-3707},
doi = {https://doi.org/10.1016/j.jog.2004.07.008},
url = {https://www.sciencedirect.com/science/article/pii/S0264370704000845},
author = {Shuzo Takemoto and Akito Araya and Junpei Akamatsu and Wataru Morii and Hideo Momose and Masatake Ohashi and Ichiro Kawasaki and Toshihiro Higashi and Yoichi Fukuda and Shinji Miyoki and Takashi Uchiyama and Daisuke Tatsumi and Hideo Hanada and Isao Naito and Souichi Telada and Nobuo Ichikawa and Kensuke Onoue and Yasuo Wada}
}

@article{tepp2024,
    author = {Tepp, Gabrielle and Stubailo, Igor and Kohler, Monica and Guy, Richard and Bozorgnia, Yousef},
    title = {Shake to the Beat: Exploring the Seismic Signals and Stadium Response of Concerts and Music Fans},
    journal = {Seismological Research Letters},
    volume = {95},
    number = {4},
    pages = {2179-2194},
    year = {2024},
    month = {03},
    issn = {0895-0695},
    doi = {10.1785/0220230385},
    url = {https://doi.org/10.1785/0220230385},
    eprint = {https://pubs.geoscienceworld.org/ssa/srl/article-pdf/95/4/2179/6524887/srl-2023385.1.pdf},
}

@article{tibet2025,
    author = {Ralte, Zosangliana and Baruah, Santanu and Gogoi, Kimlina and Tlau, Lalruatpuia and Sailo, Saitluanga and Phukan, Manoj Kumar and D’Amico, Sebastiano and Dasgupta, Sujit and Saikia, Sowrav and Bhattacharyya, Sanjeev Kr},
    title = {Seismotectonic analysis of the January 7, 2025, Mw 7.1 earthquake in Southern Tibet},
    journal = {Journal of Seismology},
    year = {2025}
}

@article{black2001introduction,
  title={An introduction to Pound--Drever--Hall laser frequency stabilization},
  author={Black, Eric D},
  journal={American journal of physics},
  volume={69},
  number={1},
  pages={79--87},
  year={2001},
  publisher={American Association of Physics Teachers}
}

@article{drever1983laser,
  title={Laser phase and frequency stabilization using an optical resonator},
  author={Drever, RWP and Hall, John L and Kowalski, FV and Hough, J\_ and Ford, GM and Munley, AJ and Ward, H},
  journal={Applied Physics B},
  volume={31},
  number={2},
  pages={97--105},
  year={1983},
  publisher={Springer}
}

@misc{spector2025demonstrationinterferometrictechniquemeasuring,
      title={Demonstration of an interferometric technique for measuring vacuum magnetic birefringence with an optical cavity}, 
      author={Aaron D. Spector and Todd Kozlowski and Laura Roberts},
      year={2025},
      eprint={2510.14064},
      archivePrefix={arXiv},
      primaryClass={physics.optics},
      url={https://arxiv.org/abs/2510.14064}, 
}

@article{Dunn2026,
author = {Eleanor Dunn and Joseph Roche},
title = {Are you ready for it? Harnessing celebrity influence for science communication and seismology – The Taylor Swift effect},
journal = {International Journal of Science Education, Part B},
volume = {16},
number = {1},
pages = {91--116},
year = {2026},
publisher = {Routledge},
doi = {10.1080/21548455.2025.2534042},
URL = {https://doi.org/10.1080/21548455.2025.2534042},
eprint = {https://doi.org/10.1080/21548455.2025.2534042}
}

@misc{ETERNA,
    author       = {Wenzel, Hans-Georg},
    editor       = {Forbriger, Thomas and Wziontek, Hartmut and Zürn, Walter and Schroth, Eva},
    year         = {2022},
    title        = {Eterna - Programs for tidal analysis and prediction},
    doi          = {10.35097/746}
}

@misc{SPOTL,
    author = {Agnew, D.C.},
    year = {2012},
    title = {SPOTL: Some Programs for Ocean-Tide Loading},
    URL = {https://escholarship.org/uc/item/954322pg},
    note = {UC San Diego: Scripps Institution of Oceanography}
}

@book{seismology,
    author = {Keiiti Aki and Paul Richards},
    title = {Quantitative Seismology, 2nd Edition},
    publisher = {University Science Books},
    year = {2002}
}

\end{document}